\documentclass[aps,prl,preprint,superscriptaddress]{revtex4-1}

\usepackage{graphicx}
\usepackage{amsmath,amsfonts,amssymb}

\begin{document}

\title{Ballistic interferences in suspended graphene}



\author{Peter Rickhaus}

\author{Romain Maurand}
\email{romain.maurand@unibas.ch}
\affiliation{Department of Physics, University of Basel, Klingelbergstrasse 82, CH-4056 Basel, Switzerland}
\author{Ming-Hao Liu}
\affiliation{Institut f\"ur Theoretische Physik,Universit\"at Regensburg, D-93040 Regensburg, Germany}
\author{Markus Weiss}

\affiliation{Department of Physics, University of Basel, Klingelbergstrasse 82, CH-4056 Basel, Switzerland}

\author{Klaus Richter}
\affiliation{Institut f\"ur Theoretische Physik,Universit\"at Regensburg, D-93040 Regensburg, Germany}

\author{Christian Sch\"onenberger}
\affiliation{Department of Physics, University of Basel, Klingelbergstrasse 82, CH-4056 Basel, Switzerland}

\date{\today}

\maketitle


\textbf{
  Graphene is a 2-dimensional (2D) carbon allotrope with the
  atoms arranged in a honeycomb lattice~\cite{novoselov_2004}.
  The low-energy electronic excitations in this 2D crystal are described by massless
  Dirac fermions that have a linear dispersion relation similar to photons~\cite{novoselov_2005, peres_2010}.
  Taking advantage of this ``optics-like'' electron dynamics, generic optical elements like lenses, beam splitters and
  wave guides have been proposed for electrons in engineered ballistic graphene~\cite{cheianov_2007, zhang_2009}.
  Tuning of these elements relies on the ability to adjust the carrier concentration in defined areas,
  including the possibility to create bipolar regions of opposite charge (\textit{p-n} regions).
  However, the combination of ballistic transport and complex electrostatic gating remains challenging.
  Here, we report on the fabrication and characterization of fully suspended graphene \textit{p-n} junctions.
  By local electrostatic gating, resonant cavities can be defined, leading to complex Fabry-P\'erot interference patterns
  in the unipolar and the bipolar regime. The amplitude of the observed conductance oscillations accounts for
  quantum interference of electrons that propagate ballistically over long distances exceeding \mbox{$1$\,$\mu$m}.
  We also demonstrate that the visibility of the interference pattern is enhanced by Klein collimation at
  the \textit{p-n} interface~\cite{young_2009, katsnelson_2006}. This finding paves the way to more complex
  gate-controlled ballistic graphene devices and brings electron optics in graphene closer to reality.}\\


\begin{figure}[htbp]
    \centering
      \includegraphics[width=1\textwidth]{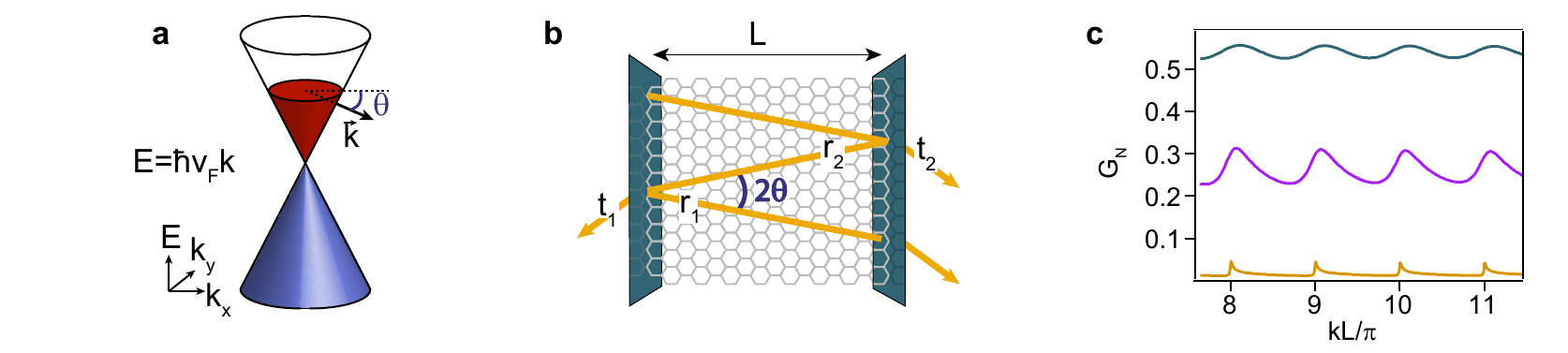}
    \caption{
      \textbf{a}, Low-energy dispersion in pristine graphene.
      \textbf{b}, Sketch of an idealized graphene Fabry-P\'erot resonator.
      \textbf{c}, Normalized conductance through an ideal graphene
        cavity as a function of the dimensionless parameter $k_{\rm F}L/\pi$ for three different
        finesses: $F=1600$ in yellow, $F=4.93$ in purple and $F=0.39$ in blue.
    }
    \label{fig:figure1}
\end{figure}


The wave nature of any physical entity, be it light or electrons, is demonstrated by measuring an interference
pattern in an apparatus, called interferometer. We distinguish two types of interferometers:
two path and multi-pass interferometers. Examples for the former are the Michelson and the Mach-Zehnder interferometer,
which are both widely used in optical experiments. In contrast, the Fabry-P\'erot (FP) interferometer is of the
multi-pass type. In optics an FP interferometer is an element that consists of a transparent glass plate
with two partially reflective surfaces on opposite sides. It can therefore be viewed as a cavity. Light within the cavity
is bouncing back and forth between the mirrors, but at each reflection a fraction of light is coupled out.
The superposition of all outgoing waves gives rise to an intensity modulation
(the interference pattern) that depends on the wavelength of the light and the distance between the mirrors.
In solid-state physics a two-dimensional conducting sheet between two electrodes
could realize an electronic FP interferometer. However, the mean free path (mfp) and phase-coherence length are usually
much shorter than the device size leading to purely diffusive incoherent electronic transport. Before the advent of 1D and 2D carbon allotropes,
long mfps and optic-like experiments have only been realized in complex semiconductor hetero\-structures~\cite{vanhouten_1989, heiblum_2003}.
%
%
During the last decade the long mfp (several hundreds of nm) accessible in low-dimensional carbon materials enabled one to explore different
ballistic phenomena. In particular, FP interferences have been demonstrated in carbon nano\-tubes~\cite{liang_2001} and
graphene nano\-structures~\cite{young_2009}.
In the former case, interferences were observed over a distance as long
as \mbox{$0.5$\,$\mu$m}, whereas the cavity size was less than \mbox{$100$\,nm} in the latter.
Nowadays, graphene junctions show a mfp close to $1 \rm\mu m$~\cite{du_2008} and electron guiding~\cite{williams_2011} as well
as transverse magnetic focusing have been revealed~\cite{taychatanapat_2013}. However, even if graphene appears
as a perfect platform for electron optics~\cite{cheianov_2007,zhang_2009}, quantum interferences with long mfp
$> 1 \rm\mu m$ and complex electrostatic gating still remain to be demonstrated.

In this letter we report FP interferences in a gate-tunable graphene \textit{p-n} junction with coherent ballistic electron motion.
We observe an interference pattern in both the unipolar and bipolar regime and for all
combinations of cavity mirrors, either defined at the source and drain contacts or at a \textit{p-n} junction in the middle of the device.


In graphene, charge carriers with small quasi-momentum $\hbar\bf k$ counted from the corner of the Brillouin zone have a linear
dispersion. In the conduction and valence band the energy $E(\textbf{k})$ as a function of wave\-vector $\textbf{k}$ is given by
$E_c(\textbf{k}) = \hbar v_F k$ and $E_v(\textbf{k})= -\hbar v_F k$, respectively, where $k=|\textbf{k}|$ and $v_F$ is the
Fermi velocity; see Fig.~\ref{fig:figure1}a. These dispersion relations yield an electron number density $n$ given by $n=\mathrm{sgn} (E) k^2/\pi$.
Taking a ballistic graphene cavity, as sketched in Fig.~\ref{fig:figure1}b, the phase difference between two
successive outgoing rays is $\Delta \phi = 2 k L \cos(\theta)$, where $L$ is the size of the cavity and $\theta$
the angle of incidence of the electron waves relative to the normal of the mirror. Summing over all partial outgoings waves, one obtains
for the transmission probability~\cite{hernandez_1986}:
\begin{equation}
T(\theta) = \frac{1}{1+F\sin^2(\Delta \phi/2)} \;  \mbox{\text ,}
\label{eq:eq1}
\end{equation}
where $F={4|r_1||r_2|}/{|t_1|^2|t_2|^2}$ with $|r_{1,2}|$ and $|t_{1,2}|$ the reflection and transmission amplitudes at the two
interfaces. $F$ is a measure of the quality factor and known as the finesse. The finesse of an
FP interferometer determines the visibility of the interference pattern, i.e., the difference in $T$ between constructive
and destructive interference relative to the mean transmission probability. The visibility is very close to unity if
$F \gg 1$ and it is small and given by $F$ in the opposite limit. Since the electric measurement of the graphene conductance
cannot distinguish the direction of the waves within the cavity, one has to integrate over the angles to deduce the expected
visibility for the conductance $(G_{\rm max}-G_{\rm min})/(G_{\rm max}+G_{\rm min})$. The normalized dimensionless conductance $G_N = (1/\pi)\int T(\theta)\cos(\theta)d\theta$
is shown in Fig.~\ref{fig:figure1}c as a function of the dimensionless parameter $kL/\pi$ for three cases
($r_1=r_2=0.8$, $F=1600$), ($r_1=r_2=0.4$, $F=4.9$) and ($r_1=r_2=0.2$, $F=0.4$). For all three cases resonances appear equidistantly whenever $k_{\rm F}L$ is a multiple of $\pi$, indicating that the conductance peaks
are mainly due to electrons at small incident angles (see Supplementary Material for details).



\begin{figure}[htbp]
    \centering
        \includegraphics[width=1.00\textwidth]{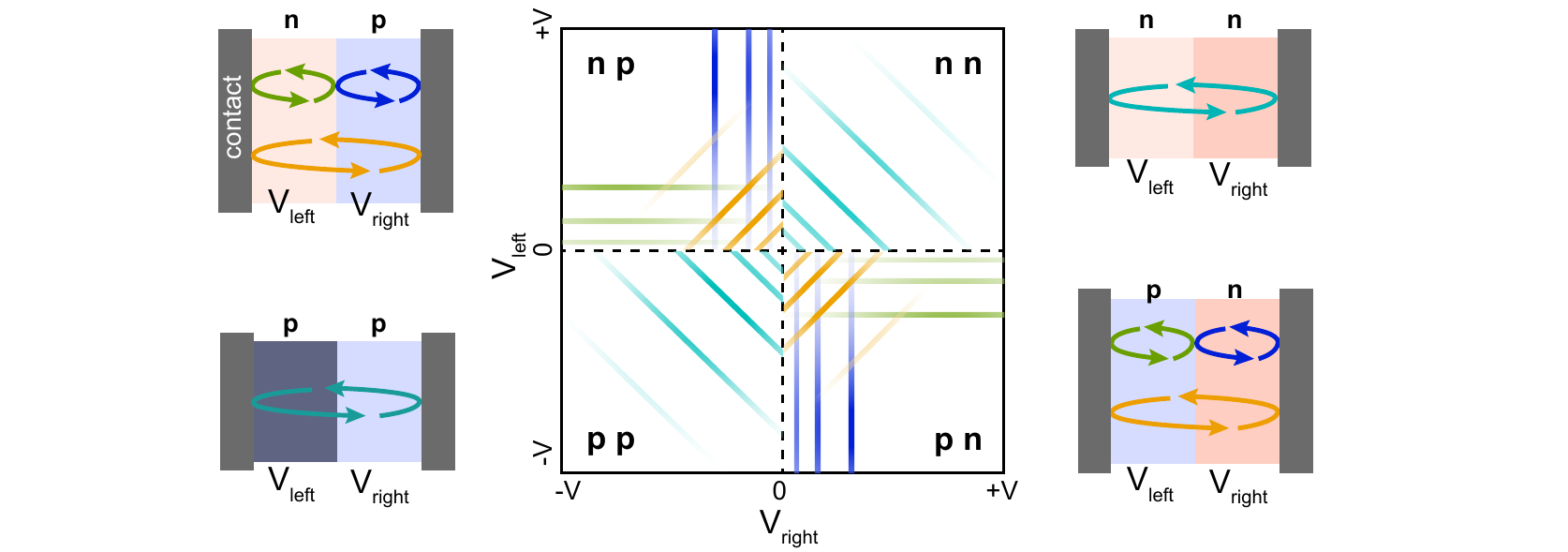}
    \caption{
      Schematic representation of all possible resonance conditions (central panel) in a FP resonator that consists of two segments of equal size (left and right panels). The carrier density and carrier type (\emph{n} or \emph{p}) in both segments can be tuned independently by $V_{\rm left}$
      and $V_{\rm right}$. FP resonances may occur due to reflection at the outer contacts or at an internal \emph{n-p} (\emph{p-n}) interface between the
      two segments. The possible resonant conditions are indicated by color-coded lines for the respective cases in the middle part.
      }
    \label{fig:figure2}
\end{figure}

We now focus on a graphene FP cavity that consists of two segments: left and right.
 Assume that in both segments the carrier density and its sign (\emph{n} and \emph{p}) can be controlled independently by two gate voltages $V_{\rm left}$
and $V_{\rm right}$ with identical efficiency. Figure~\ref{fig:figure2} shows in a schematic 2D conductance map the expected FP interference patterns as a function of $V_{\rm left}$ and $V_{\rm right}$. This map has four
quadrants corresponding to the four polarity configurations~\cite{williams_2007}. If the left and right segments have
the same polarity (\emph{nn} or \emph{pp}), the cavity is in the unipolar regime.
In contrast, in the case of opposite polarities (\emph{np} or \emph{pn}), the cavity is in the bipolar regime.
In the unipolar regime only interferences of electrons bouncing back and forth between the two outer contacts are expected (lower left and upper right panel in Fig.~\ref{fig:figure2}).
The corresponding beating pattern evolves along $V_{\rm left}=V_{\rm right}$ and is indicated by the light-blue
lines in the central panel of Fig.~\ref{fig:figure2}.
In the bipolar case, assuming a semi-transparent \textit{p-n} interface between the segments,
the junction can be considered as two cavities in series, each controlled by its respective gate.
Resonances in the left (right) cavity depend only on $V_{\rm left}$ ($V_{\rm right}$) and are represented by green (blue) lines.
Another distinct set of resonances in the bipolar regime can arise from charge carriers that tunnel through the central \textit{p-n} interface in the middle
and bounce back and forth between the outer contacts. The corresponding beating pattern should evolve along the condition $k_{\rm left} \approx k_{\rm right}$. In terms of gate voltages this condition corresponds to $V_{\rm left} \approx -V_{\rm right}$ as indicated in Fig.~\ref{fig:figure2} by the orange lines.


\begin{figure}[htbp]
    \centering
    \includegraphics[width=1.00\textwidth]{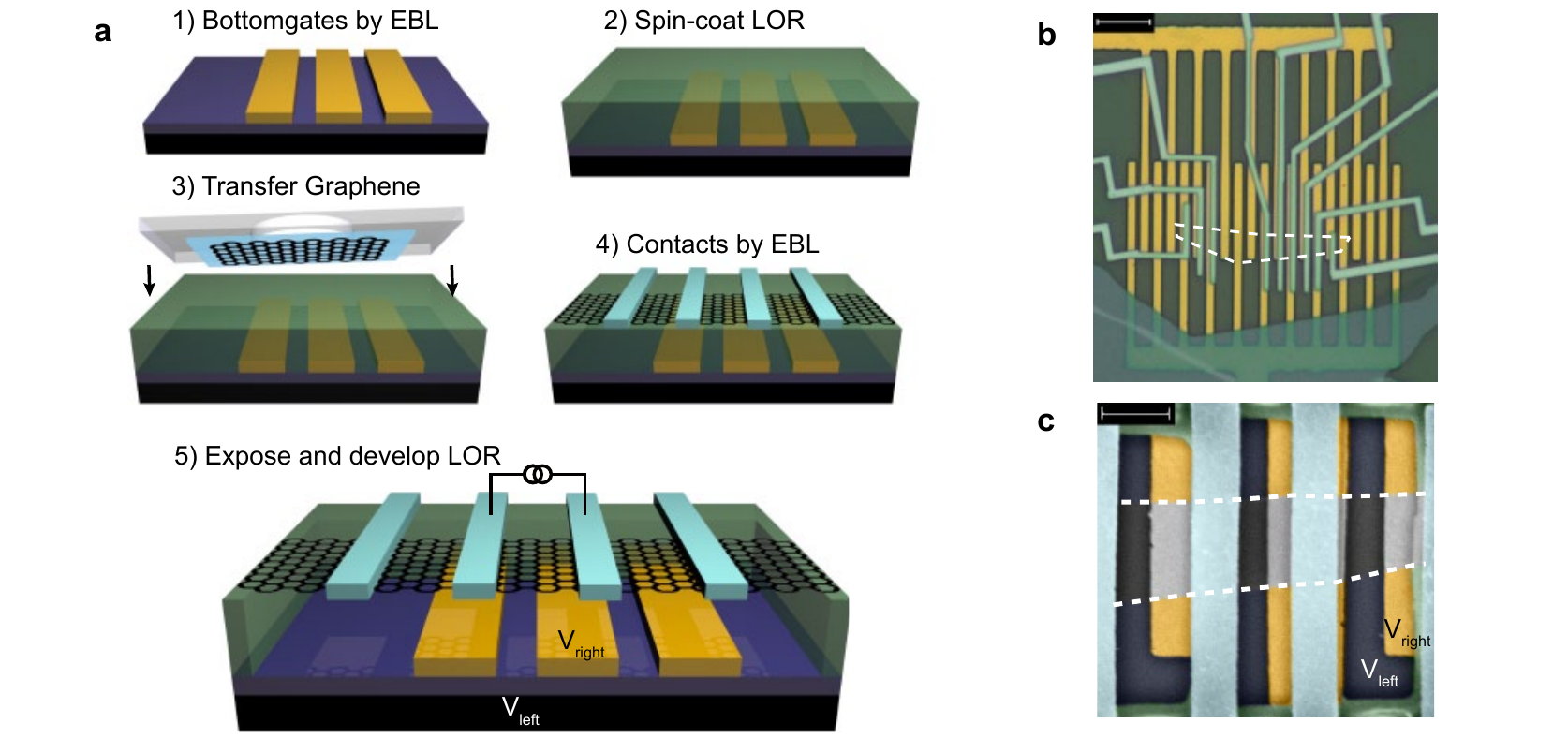}
    \caption{
      \textbf{a}, 3-dimensional illustration of the device fabrication consisting of the following steps:
        1) bottom gates are realized by standard e-beam lithography on a doped $\rm Si$ wafer with $300$ nm thick $\rm SiO_2$ top layer and
        2) covered by a $700$ nm thick LOR (lift-off resist) layer.
        3) Graphene is exfoliated on a separated wafer covered by a stack of PVA/PMMA, then aligned to the bottom gates and transferred.
        4) Contacts to the graphene are then realized with standard e-beam lithography followed by an evaporation of $50$ nm palladium.
        5) Finally, the device is suspended by selectively exposing the LOR with a large dose and developing it.
      \textbf{b}, Optical image of several junctions defined within the same graphene flake indicated by white dashed lines.
        Bottom gates are in yellow and ohmic contacts in blue. The LOR resist appears in dark green. Scale bar, $5$ $\mu$m.
      \textbf{c}, Scanning-electron microscopy image of three graphene \textit{p-n} junctions in a row.
        Bottom gates are shown in yellow and contacts in light blue. Scale bar: $1.2$ $\mu$m. }


    \label{fig:figure3}
\end{figure}

To realize experimentally such a tunable FP cavity, devices were prepared by combining a mechanical transfer process~\cite{dean_2010}
with a hydrofluoric acid-free suspension method proposed by Tombros~{\it et al.}~\cite{tombros_2013}. The fabrication process is schematically
shown in Fig.~\ref{fig:figure3}a. For a complete description, see \textbf{Methods}.
Clean graphene is obtained by an \textit{in-situ} current annealing process. Figure~\ref{fig:figure3}b shows a colored optical image of several
junctions realized within the same graphene flake indicated by white dashed lines. An SEM micrograph of three \textit{p-n} junctions
is presented in Fig.~\ref{fig:figure3}c. A voltage applied on the bottom gate ($V_{\rm right}$, yellow electrodes in
Fig.~\ref{fig:figure3}b,c) will tune the charge carrier density on the right side of a junction, while a voltage applied on the back gate
($V_{\rm left}$) will act on the left side. The device was measured in a $^4\rm He$ cryostat at a temperature of $T=1.5$\,K.
Differential conductance, $G=dI/dV$ was measured by standard lock-in techniques with a voltage bias of $100$ $\mu$V at  $77$ Hz.
Series and contact resistances were not subtracted from the data.


\begin{figure}[htbp]
    \centering
    \includegraphics[width=1.00\textwidth]{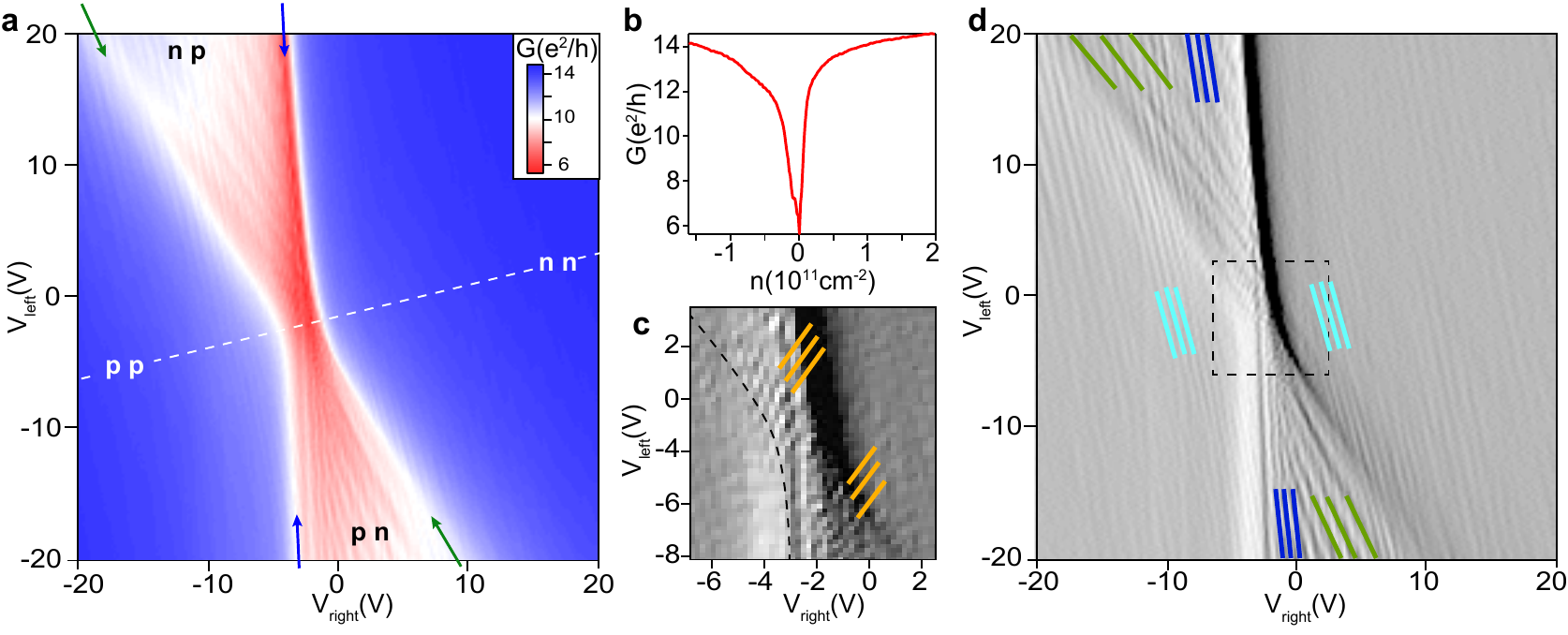}
    \caption{
      \textbf{a}, Two terminal differential conductance $G=dI/dV$ as a function of $V_{\rm right}$ and $V_{\rm left}$ at $T=1.5$ K,
        demonstrating independent control of carrier type and density in the left and the right side of the graphene sheet.
        Labels in each of the four conduction regimes indicate the carrier type.
      \textbf{b}, Conductance as a function of charge carrier density $n$ along the dashed white line of \textbf{a}.
      \textbf{c}, Zoom into the transconductance map presented, in \textbf{d}, around the charge neutrality point.
      \textbf{d}, Transconductance obtained from the numerical derivative with respect to $V_{\rm right}$ of \textbf{a} emphasizing conductance oscillations.
        Color-coded lines link the observed conductance oscillations with the FP interference pattern expected from the simple picture of
        Fig.~\ref{fig:figure2}: light blue lines in the unipolar regime, green and blue lines for interferences on either side of the graphene in the bipolar case and orange lines for full length interference in the bipolar regime.}
    \label{fig:figure4}
\end{figure}

Figure~\ref{fig:figure4} shows the main experimental result of this work. In part a) the differential conductance between the two contacts as a function of $V_{\rm right}$ and
$V_{\rm left}$ is presented for a junction of length $L=1.2$~$\mu$m and
width $W=3.2$~$\mu$m. As expected, this 2D plot reveals four regions indicated by the labels
\emph{pp}, \emph{nn}, \emph{np} and \emph{pn} corresponding to different carrier types in the two sides of the sample. The borders
between the unipolar and bipolar regions coincide with the
charge neutrality point of the left and right graphene segments (indicated by blue and green arrows respectively).
The border lines between the different regions are not perpendicular to each other, as expected for independent gates and discussed above in Fig.~\ref{fig:figure2} for two reasons. First, both the back gate and the bottom gate are at a distance to the graphene sheet that is comparable
to the length of the device. This leads to a cross coupling between the two gates i.e., the right gate also affects the left graphene segment and \emph{vice versa}.
Second, the back gate (left gate) is below the bottom gates and therefore strongly screened by these, leading to a much weaker gate coupling from the back gate to the graphene than from the bottom gate.

A slice through the 2D conductance plot along the white dashed line is shown in Fig.~\ref{fig:figure4}b as a function of
carrier density $n$ which we estimated from a parallel plate capacitor model. A positive sign refers to the \emph{n}-region and a negative sign likewise to the \emph{p}-region. A change in gate-voltage of $1$V along the white
dashed line induces a change in density of $n\approx 9\times 10^9$~cm$^{-2}$. The graph exhibits a remarkably sharp conductance dip, corresponding to the charge neutrality in the entire graphene sheet. The fact that this dip also occurs close to zero gate-voltage reflects the high quality of the graphene sample.
The conductance saturation and the asymmetry at high carrier density suggest an \emph{n}-type doping of the graphene
contacts below the palladium contacts~\cite{huard_2008,giovannetti_2008}.
By defining the mobility as $\mu =1/e \times d\sigma/dn$ with $\sigma$ the graphene conductivity, we estimated a mobility of
$\mu \approx 150\times 10^3$~cm$^2$/Vs at $n=4\times 10^9$~cm$^{-2}$. The junction reaches a minimal conductivity of
$\sigma_{\rm min} \approx {2e^2}/{h}$ close to the ballistic limit of ${4e^2}/{\pi h}$~\cite{tworzydlo_2006} indicating a weakly disordered graphene sheet~\cite{mayorov_2012, du_2008}.

The most striking features in the conductance map are, however, the oscillation patterns visible in both the unipolar and bipolar configuration. To emphasize these features, the transconductance $dG/dV_{\rm right}$ numerically calculated from the data in Fig.~\ref{fig:figure4}a is presented in Fig.~\ref{fig:figure4}d. The different interference patterns expected from the simple picture of Fig.~\ref{fig:figure2} are well revealed in this map. In the unipolar case, one global interference pattern is visible (highlighted in a small region by light-blue lines) corresponding
to FP interferences of electron waves that bounce back and forth between the outer metallic contacts.
In the data the density separation $\Delta n\approx 10^{10}$~cm$^{-2}$ between constructive interference peaks at around $n=10^{11}$~cm$^{-2}$. Based on the equation $n=\mathrm{sgn}(E) k^2/\pi$ and the resonance condition $\Delta k L_{cavity}=\pi$ we derive
the relation $\Delta n = 2\sqrt{\pi n}/L_{\rm cavity}$. This yields an effective cavity length $L_{\rm cavity} \approx 1.1$~$\mu$m
in good agreement with the geometrical length $L=1.2$~$\mu$m of the junction . In the bipolar regime at high doping two clear
oscillation patterns are visible. One pattern is mainly tuned by $V_{\rm right}$, indicating that resonances arise from the cavity
created at the right side of the \textit{p-n} interface. This pattern is highlighted by blue lines in Fig.~\ref{fig:figure4}d.
The other pattern marked by green lines is predominantly tuned by $V_{\rm left}$, indicating that it originates from the cavity formed at the left side of the \textit{p-n} interface.
The area in the transconductance map marked by the dotted square is depicted enlarged in Fig.~\ref{fig:figure4}c. It shows yet another oscillation pattern (orange lines). It follows gates values such that $n_{\rm left} \approx -n_{\rm right}$ with an interference period $\Delta n \approx 3\cdot 10^{9}$~cm$^{-2}$ at $n = 10^{10}$~cm$^{-2}$ leading to
$L_{\rm cavity}\approx 1.2$~$\mu m$. We conclude that these oscillations must arise from quantum interference over the full device length, i.e. charge carriers that tunnel through the \textit{p-n} interface, 
 changing their character from \emph{n} to \emph{p} and back.

We emphasize at this point that there is no evidence of disorder in the transconductance plot of Fig.~\ref{fig:figure4}d. All
patterns are regular and can be ascribed to distinct FP cavities, some of which extend in length over the whole device. The occurrence of FP interference patterns implies that the phase-coherence length exceeds twice the system size. Random disorder would then generate further interference patterns, but these are never regular. The fact that we can deduce from the interference pattern effective cavity-lengths from contact to contact and that we do not see any irregular features are strong signs for ballistic coherent transport over distances exceeding $2$~$\mu$m.


\begin{figure}[htbp]
  \centering
  \includegraphics[width=1.00\textwidth]{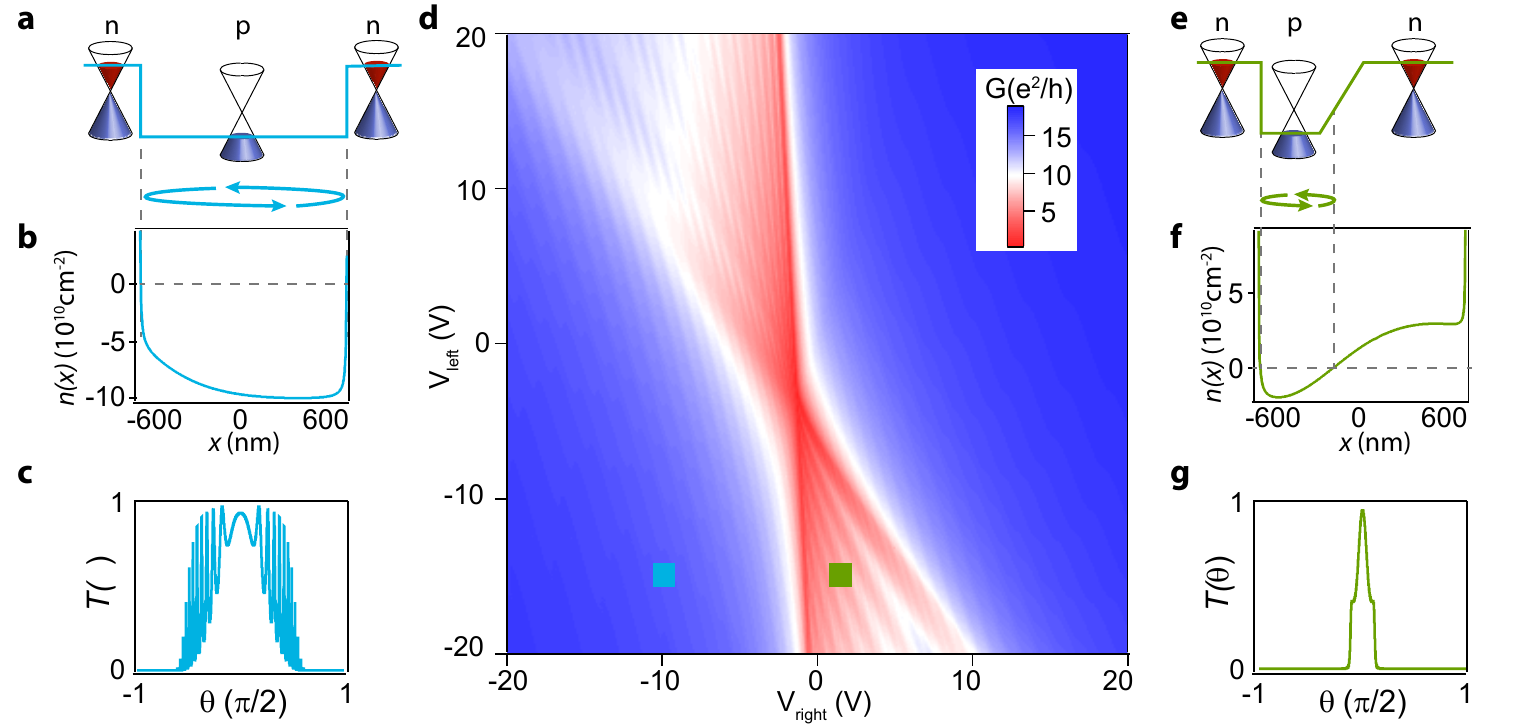}
  \caption{
    \textbf{a}, Schematic view of charge carrier configuration at the unipolar point $(V_{\rm left},V_{\rm right})= (-15,-10)$V
      indicated by a blue square in \textbf{d}.
    \textbf{b}, Calculated charge carrier profile at this point, and
    \textbf{c}, associated simulated transmission function \textit{versus} incident angle $\theta$.
    \textbf{d}, 2D conductance map obtained by tight-binding simulation as a function  of $V_{\rm left}$ and $V_{\rm right}$.
    \textbf{e,f,g}, Same as \textbf{a,b,c} but for parameters at the bipolar point $(V_{\rm left},V_{\rm right})= (-15,2)$V indicated
      by a green square in \textbf{d}.}
  \label{fig:figure5}
\end{figure}

To compare the experiment with theory, the charge transport through the graphene device was numerically calculated in the phase\-coherent ballistic regime using the actual geometrical parameters
of the device (see Supplementary Material). The computed conductance is shown in Fig.~\ref{fig:figure5}d, which presents a purely ballistic tight-binding transport simulation using a real-space Green's function method within the Landauer-B\"uttiker formulation ~\cite{liu_2012} combined with the quantum capacitance model for carrier density computation~\cite{liu2_2012}. Since no dephasing or energy averaging has been incorporated, the
obtained conductance modulations show the maximal possible visibility for this device geometry.
The correspondence between theory and experiment is remarkably good and supports the picture of ballistic motion drawn from experiment.

As pointed out before, we do see all the expected FP resonance patterns due to the formation of distinct cavities having
different doping and different types of mirrors. However, the strength of the oscillation pattern (the visibility) varies
in different regions. This is linked to the reflection properties of our cavity "mirrors", determining the cavity finesse.
Cavities can be created either by  sharp or smooth unipolar or bipolar potential steps.

Let us first consider a unipolar potential step. If $d$ is the length over which the carrier density varies and $k_F$ the mean Fermi wave\-vector, the junction
is assumed smooth if $k_F d \gg 1$ and sharp in the opposite limit. In the smooth case the electrons follow adiabatically
the potential evolution. They may bend similar to light rays in a medium with a slowly varying refractive index, but they will not
back\-scatter. In contrast, for a sharp potential step one might expect back\-scattering. However, charge carriers in graphene are
chiral, i.e. they have a pseudo\-spin that is intimately tied to the wave\-vector. Scattering from a state with wave\-vector $\bf k$ into
$\bf - k$ is forbidden, because this would flip the pseudo\-spin. The conservation of pseudo\-spin is only lifted,
if the potential step is abrupt on the atomic scale \cite{cheianov_2006}, which is not the case here. Thus pseudo\-spin conservation
implies that effectively all unipolar potential steps are highly transmissive for almost every incident angle~\cite{cayssol_2009}
(only electrons with incident angles close to $\pi/2$ have an appreciable backscattering probability). Cavities
created by unipolar interfaces have consequently a small finesse and will display small visibilities.

We turn now to the reflection properties of a \emph{p-n} junction. We assume that the carrier densities on both sides
are of opposite sign but equal in magnitude and that $k_F$ denotes the wave\-vector in the homogeneous region outside the
junction. In the case of a sharp \emph{p-n} interface with $k_{\rm F}d \ll 1$ the transmission probability is given
by $T(\theta) = \cos^2(\theta)$~\cite{cheianov_2006}. The corresponding cavities have a small finesse for
electron waves over a large range of incident angles around $\theta = 0$ and will therefore have a small visibility.
The situation is different for a smooth \emph{p-n} junction. Since the charge carrier density evolves smoothly through zero,
essentially all electron trajectories are adiabatically refracted off the interface (total reflection) except for the normal one, which
can Klein-tunnel through the junction~\cite{cheianov_2006, katsnelson_2006}.
The transmission probability through such a smooth \emph{p-n} junction is given by the expression $T(\theta)=\exp(-\pi k_{\rm F}d \sin^2\theta)$~\cite{cheianov_2006}
which also predicts perfect transmission at normal incidence. Similar to the sharp interface one could expect the visibility of the interference pattern to be small. However, Klein tunneling does not only yield full transmission at normal incidence
but also leads to a strong collimation in transmission. Therefore, almost all trajectories that are incident under a small angle ($\theta \not = 0$) are
exponentially suppressed and do not add to the total conductance. This suppression of the background current increases
the overall visibility. Although the finesse vanishes for $\theta=0$
the contribution from small angles (with $\theta \not = 0$) yield a higher visibility than is the case for unipolar or sharp bipolar
cavities (see Supplementary Material for further details).

Based on the preceding discussion, we now compare the observed visibilities for two particular cases in a) the unipolar and b) bipolar regime
at gate voltages $(V_{\rm left},V_{\rm right})= (-15,-10)$V and $(-15,2)$V, respectively. These two points are marked in
Fig.~\ref{fig:figure5}d by light-blue and green squares. For both cases, the reflection properties at the contacts are important
parameters and need to be known. The full comparison of the simulation with the experiment reveals that the contacts dope graphene as \emph{n}-type.
In addition to this doping the simulation includes a small mass term in the contacted region~\cite{kwon_2009}. For the unipolar case, the
simulation yields the charge carrier density profile plotted in Fig.~\ref{fig:figure5}b. In this configuration the interior of the FP cavity is
\emph{p}-type with mirrors created by the \emph{p-n} interfaces close to the Pd contacts. The latter can be considered to be relatively
sharp due to the rapid decay of the contact-induced screening potential~\cite{khomyakov_2010}. Consequently, these interfaces are almost transparent and give
rise to low finesse for the FP interferences as found in the experimental and the numerical conductance map with respective average
visibilities of $0.5\%$ and $2\%$.
In the bipolar case the simulation yields the carrier density profile plotted in Fig.~\ref{fig:figure5}f.
It presents an extremely smooth \emph{p-n} interface created electrostatically in the middle of the graphene sheet.
The two cavities created on the right and left side will present a much higher finesse than in the unipolar configuration
which consistently leads to higher visibilities in both the experimental and numerical conductance maps.
The average visibility for the experiment is around $5\%$, whereas the simulation yields $20\%$.
The calculated angle dependent transmission probabilities for both regimes are plotted in
Figs.~\ref{fig:figure5}c and g respectively, which clearly demonstrate that the visibility difference between the unipolar and bipolar
regime is a direct consequence of the exponential collimation of Klein tunneling~\cite{katsnelson_2006, cheianov_2006,cayssol_2009}.


In conclusion, we have fabricated and characterized a suspended
graphene \textit{p-n} junction. Fabry-P\'erot type conductance oscillation patterns
visible in both unipolar and bipolar regimes point to an
extremely long mean free path and phase-coherence length $> 2$~$\mu$m accessible in this device. Moreover we have shown that the visibility
difference between the unipolar and the bipolar regime is due to Klein collimation occurring at a
smooth \emph{p-n} interface in graphene.
Our experiment paves the way to more complex engineered ballistic graphene to realize electron
lensing, focusing~\cite{cheianov_2007} and efficient guiding~\cite{zhang_2009,williams_2011}.

\section{Acknowledgements}
\begin{acknowledgments}
We are indebted to A. van der Torren for the development of the transfer technic.
This work was financed by the Swiss NSF, the ESF programme Eurographene, the EU FP7 project SE$^2$ND, the ERC Advanced Investigator Grant QUEST, the Swiss NCCR Nano and QSIT.
M.-H.L. was supported by Alexander von Humboldt foundation. He and K.R. acknowledge the financial support of Deutsche Forschungsgemeinschaft (SFB 689).
\end{acknowledgments}

\section{Authors contributions}

P.R. and R.M. contributed equally to this work. P.R. and R.M. fabricated the devices. Measurements were performed by P.R., R.M., M.W.. M.H.L. performed the conductance simulations and calculations of carrier densities. C.S. guided the experimental part of the work. K.R. guided the theory part of the work. All authors worked on the manuscript.

\section{Methods}
\label{sec:methods}

Suspended graphene \textit{p-n} junctions were obtained by combining a mechanical transfer technique with a hydro\-fluoridric acid free suspension method developed by Tombros~{\it et al.}~\cite{tombros_2013}.
The fabrication starts with the spinning of a polymer stack consisting of a water-soluble PVA and PMMA layer
on an oxidized ($300$~nm) Si wafer, on which alignment markers are predefined.
Graphene sheets are then mechanically exfoliated on this stack and located by optical means.
Depending on the location of the selected graphene flake, $5/45$~nm thick Ti/Au bottom gate electrodes are
tailor-made on a separate oxidized Si wafer by standard e-beam lithography and covered by a $1\mu$m thick
LOR resist, Fig.~\ref{fig:figure3}a-1,2. Afterwards, the wafer that contains the graphene is immersed in
deionized water. Once PVA is dissolved, the wafer sinks to the bottom, leaving the PMMA-graphene membrane floating on top at the
liquid-air interface. Next, this membrane is fished out by a transfer glass, dried and fixed in place of the
UV-mask in a modified MJB3 mask aligner. The other wafer, containing the bottom gates and LOR layer, is fixed
at the wafer side of the mask aligner. The graphene flake is then carefully aligned to the bottom gates and finally
transferred in contact by heating the stack to $120\,^{\circ}\mathrm{C}$, Fig.~\ref{fig:figure3}a-3.
Once transferred, PMMA is dissolved in xylene. Palladium contacts ($50$~nm in thickness) are then fabricated through PMMA e-beam
lithography with xylene as developer and lift-off solvent, Fig.~\ref{fig:figure3}a-4 and optical image in Fig.~\ref{fig:figure3}b.
In the final step the LOR layer below the graphene is e-beam exposed and developed in ethyl-lactate to obtain suspended
graphene junctions, Fig.~\ref{fig:figure3}a-5. After mounting the finished device in a variable temperature $^4 \rm He$ cryostat, we perform current annealing by
applying a DC current at $1.5\rm K$. In this procedure the current is increased until the conductance as a function of a gate voltage
$V_{\rm gate}$ shows a pronounced dip around $V_{\rm gate}=0\rm V$, reminiscent of a high quality device. This requires current
densities that usually reach values up to $350\mu \rm A/\mu \rm m$.

\end{document}